# Overcoming the EPR Paradox

Ghenadie Mardari

Open Worlds Research, Baltimore, USA.

**Abstract**: Quantum experiments detect particles, but they reveal information about wave properties. No matter how quanta are detected, they always express the local net state of the corresponding wave-function. The mechanism behind this process is still a mystery. However, quantum wave-functions evolve like classical waves. If they determine all the observations, why do they entail "weird" phenomena? In particular, why do they produce conceptual problems, such as the EPR paradox? It turns out that all the major interpretations in quantum mechanics are particle-based ontologies. Variables of fundamental interest, such as momentum and position, are assumed to reflect the input states of discrete entities, even when their evolution is predicted by wave equations. Here I show that a whole class of microscopic phenomena, including entanglement, can be interpreted without contradictions if the variables of quantum mechanics are actually treated as wave properties. The main recommendation is to assume that sharp spectral states are transient outcomes of superposition, rather than permanent input components.

## Introduction

The EPR paradox *(1)* is – above everything else – a logical conundrum. In a theory without contradictions, objective entities should express one type of behavior or another, but not both at the same time. Yet, quantum mechanics appears to defy this rule. When two particles are entangled, their states are perfectly correlated. If one of them is narrowed down by measurement, the other one is known to be sharp as well. If both of them are measured, but in different ways, then sharp values can be obtained for two variables at the same time, even if the latter do not commute. By implication, entangled quanta should not be able to violate Bell's inequality. And still, quantum mechanics predicts the opposite *(2,3)*. This is a true paradox. If the states are sharp, then they must commute. If they violate Bell's inequality, then they cannot commute *(4-8)*. How is it possible for the same states to be sharp and broad at the same time? As will be shown below, the source of this problem can be traced to a conceptual ambiguity, pertaining to the nature of linear superposition. Essentially, it is a mistake to assume that two sharp states belong to the same quantum at the same time. Though, a note on the relevant definitions is required before this discussion can proceed. Throughout the rest of this text, quantum properties are going to be described as "uncertain" if their statistics contain multiple separable states (acting one at a time). They are going to be described as "superposed", with either sharp or broad spectra, if their statistics contain multiple inseparable states (acting all at the same time). To avoid confusion, quanta will be described as "undefined" only if they can express incompatible types of behavior in alternative measurements. In particular, a property with broad spectrum is still considered to be "well-defined" in this context. When a variable is expected to be "sharp" *and* "broad" at the same time, then the quantum is "undefined", because incompatible histories are required for each manifestation. This will make it easier to capture the essence of the debates between Einstein and Bohr on the nature of entanglement.

To this day, the interpretation of quantum behavior is impaired by the widespread endorsement of two misconceptions. Number one is the perception that quanta change their properties for every measurement, as if they "know" which type of device is going to be used in each case. For example, elementary particles have one spectral profile if their momentum is captured sharply, and a completely different profile if their position is pinned *(9)*. This fact encourages a visual image in which different devices (a momentum detector versus a position detector) must be chosen by observers, with incompatible results. In actuality, both properties can be determined with the same type of device – an event counter (Fig.1). The *coordinates* of each detection reveal *information* about position and momentum with various levels of sharpness. Hence, the detector can be moved to a location in which the spectrum of position is narrow but the spectrum of momentum is wide, or to a location in which the opposite profile takes shape. In other words, it is the observers who behave as if they know where to place their detectors, in order to obtain sharp values for one variable or the other. Quantum properties do not change spontaneously. They always evolve as predicted by the corresponding wave-function *(10)*.

Misconception number two concerns the role of quantum uncertainty. Some variables are necessarily detected with broad spectra, which means that several states are attributed to a single particle at the same time. Yet, a single measurement can only capture one state. Therefore, it is often assumed that quanta express only one spectral component at a time, with a corresponding degree of uncertainty about their "real" state prior to the act of observation. The truth is that quanta express the inseparable effect of all the components in every single measurement *(11, 12)*. Detection events reveal the local net state of superposition, *i.e.* the sum of all the possible state vectors that overlap on a relevant point of detection. For example, quanta cannot be observed in the areas with destructive interference in the double-slit experiment. If all the components cancel out, quanta are physically absent (which can only happen if all the components are active at the same time, in order to cancel out). Instead, of course, they are found in excess in the areas with constructive interference. Accordingly, there can be no uncertainty as to "which component is real" when it comes to superposed quantum properties. Instead, there is *certainty* that all the components are involved at the same time, just as it happens at any point in the cross-section of a classical wave-front. To be clear, quantum states are uncertain prior to detection, because it is impossible to predict the coordinates of individual events. Though, any one of those probable outcomes is going to express the net effect of several superposed components.

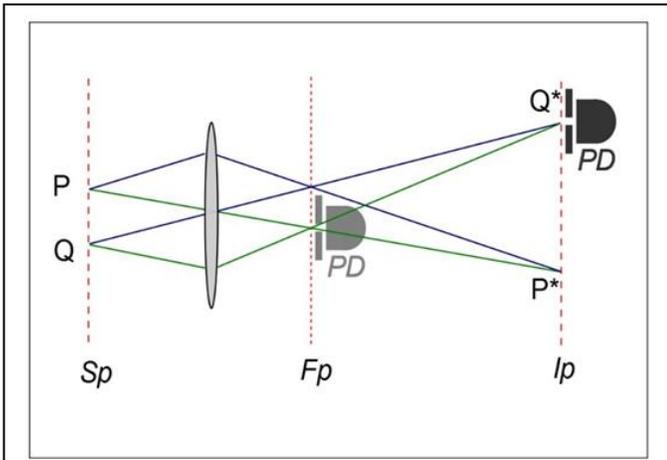

**Fig. 1. Geometrical representation of non-commuting wave properties.** Classical wave-fronts and quantum wave-functions have identical patterns of evolution in free space. After passing through a lens, the waves have sharp momentum spectra and wide position spectra in the focal plane *Fp*. They have sharp position spectra and wide momentum spectra in the image plane *Ip*. (Only two virtual components are shown, for clarity). The same pinhole detector *PD* can be moved from one plane to the other, in order to capture incompatible profiles. Hence, quantum properties are not influenced by the method of detection. Observable qualities change because the local wave-function profile changes. Caution: the rays are intended to capture the structural properties of the wave-function, not the trajectories of individual quanta.

In a nutshell, the tradition is to assume that quantum measurements examine particle properties that are created at the source. Many input states are possible, but only one is captured. (Alternatively, several components remain unresolved, because they happen to be indistinguishable). In contrast, the formalism of quantum mechanics predicts the observable net states of relevant wave-functions, as they happen to emerge at various locations. There is an obvious mismatch between these two perspectives, because particle properties have to maintain a fixed spectrum width throughout the process of propagation. Yet, the net state of the wave-function is produced by summing up component wave amplitudes, with changing phase differences. In some contexts, the same inputs can add up to a net state with narrow spectrum (because components cancel out), or to a net state with very wide spectrum. Even sharp quantum states are produced by superposition. Instead of being stripped down to a permanent input component, the wave-function creates a higher-level transient property through interference. Hence, the goal of this essay is to expose the gap between the leading interpretations and the actual content of the formalism of quantum mechanics. Specifically, many puzzles in this field emerge from the confusion between the input profile of a particle beam and the geometrical profile of a wave-function. The roots of this problem go back to "classical times", when it was expedient – and seemed harmless – to analyze wave behavior in terms of particle trajectories (*viz*., the method of ray tracing). In the final analysis, it is the interpretation (rather than the formalism) that needs to be improved as a solution for the EPR paradox.

## The problem

Quantum mechanics was called into existence by the discovery of two microscopic phenomena: the corpuscular nature of electromagnetic waves and the wave-like nature of subatomic particles. In the history of science, this was a dramatic turn of events, forcing the formulation of a new class of theories. Yet, those findings were not enough to overturn the conceptual foundations of pre-modern physics. There is a clear difference between "surprising" observations and "incomprehensible" patterns of behavior. Indeed, classical mechanics is compatible with the existence of discrete objects surrounded by fields and other types of dual phenomena. Accordingly, quantum theory was not expected to be "irreducibly weird" from the start. It became perceived like that only later, when the details of quantum interactions defied specific predictions, derived from classical models. In order to understand this problem, it is necessary to review a subtle aspect of linear wave superposition.

Mechanical waves in elastic media do not transfer matter. They represent propagating states of motion. However, it often looks *as if* groups of particles are flowing with rectilinear trajectories, when running waves are observed. The curious feature of these "virtual particles" is their apparent ability to pass through each other unperturbed. If two coherent wave-trains intersect, they do not show any trace of interaction after overlap. Whether a beam of sound is present alone, or whether it crossed a number of beams on its path, the output profile remains the same. On the other hand, the process of interference is also obvious in the volume of overlap between beams. How is it possible for the waves to interfere, and yet to behave as if they never did? The classical answer is provided by reference to the principle of superposition (*13-15*). When several waves overlap, they are assumed to remain unperturbed. Yet, they are also indistinguishable, because they act simultaneously on detecting devices. For this reason, only joint effects are observable. The rule is that the amplitude of the observable net wave is equal to the vector sum of the component amplitudes at each point. If two waves overlap crest to crest, the net state is a larger wave (constructive interference). If they overlap crest to trough, the net state is a smaller wave (destructive interference). In other words, wave behavior can always be represented by a model with virtual "running" components that go through each other unperturbed. This is an essential element of geometrical analysis, where the method of ray tracing represents the waves with clusters of rectilinear trajectories.

Electromagnetic waves are not mechanical waves. As shown by Michelson and Morley (*16*), their relative speed of propagation is invariant and cannot be reduced to a pattern of motion transfer on a medium. Moreover, the discovery of quantization (and especially the photon hypothesis) (*17*) made it appealing to imagine the propagation of optical beams as if they were streams of real particles. Indeed, the method of ray tracing yielded the highest dividends in geometrical optics, and it only seemed appropriate to suspect that optical "rays" were in fact particle trajectories. This hypothesis had a very straightforward and easily verifiable implication: quantum interference should be impossible! In classical wave mechanics, there are no means for separating the effects of individual components in superposition. Those effects are macroscopic. In quantum mechanics, corpuscular constituents of each wave can be detected one by one. Given the rule that input waves remain unperturbed during overlap, it only makes sense to suppose that elementary components of such waves should also remain unperturbed. In other words, a double-slit optical projection should have interference fringes at the classical level, but not at the quantum level. Unfortunately, this expectation was falsified conclusively by real-life experiments ([18, 19](#)). Specifically, quanta were found to be present in excess in the areas with constructive interference, while missing in the areas with destructive interference. Despite this clear evidence of perturbation, the quanta were still detected in their original ("rectilinear") paths after the volume of overlap, as if no interference ever took place.

The discovery of quantum interference undermined the classical version of the "underlying reality", at least for electromagnetic waves. The paradigm of the time entailed that interference was a measurement artifact. The actual process had to consist of particles with rectilinear trajectories. Yet, interference fringes were obtained at very small rates of emission – even when the amount of energy was sufficient for just a single quantum at a time. In other words, the so-called "measurement artifacts" persisted at the lowest practical levels of detection. By implication, the corresponding "real" states could never be seen. How can it be that something unobservable is more real than something observable? After raising this question, Niels Bohr flipped the classical paradigm upside down. He declared that "what is real" cannot be considered in isolation from "what is measured" ([20](#)). Therefore, the so-called "measurement artifacts" should be treated as real, rather than the unmeasured components. Unfortunately, this made it difficult to explain the nature of incompatible measurements with self-consistent stories. How can it be that interference is real if quanta are measured in the interference volume, and not real if measured afterwards? In order to avoid this problem, Bohr's postulate could not be treated as a mere convention. It had to be raised to the status of ontological law. In particular, observable quantum properties had to be interpreted as fundamentally real. Measurements did not obscure a well-defined underlying reality. In some cases, alternative experiments required contradictory explanations. Yet, the observable reality was never inconsistent, because the corresponding quantum measurements were mutually exclusive ([21](#)). This was the basis for Bohr's famous complementarity principle: underneath the well-ordered detected phenomena there was no independent well-defined "reality". All of the potential histories had to co-exist at the same time, until the act of measurement allowed one of them to materialize. *Contraria sunt complementa.*

Bohr's interpretation required a huge leap of faith from fellow scientists, and not all of them were willing to make it without opposition. Yet, these objections had a strange tendency to produce ironic results. Most famously, Einstein, Podolsky and Rosen ([1](#)) pointed out that entangled quanta had perfectly correlated states. If one of them was measured, then the second one was also determined. This meant that – contrary to Bohr's postulate – it was possible for unmeasured particles to have well-defined properties in the absence of direct measurement. Once a quantum was observed, its entangled partner had to be either sharp, or broad, but not both (for the same property). Unfortunately, there were no restrictions on the type of measurements that could be performed on the first particle. If it was tested in a context that revealed a sharp property, then the entangled partner was sharp. If the test exposed a broad property, then the entangled partner was broad. In other words, the spectrum of the same unmeasured property had to be objectively sharp *and* broad at the same time. Its real status could only be determined if the correlated partner was detected in one way or the other. For clarity, a broad state expresses the joint effect of several component vectors and is physically different from a sharp state. It is not possible to just blame the measurement for having "low resolution", because the property (in all of its complexity) is expressed by a single quantum. Consequently, the EPR argument appeared to strengthen Bohr's interpretation. It seemed even more compelling to assume that multiple histories were virtual at the same time, until an act of measurement forced the Universe to realize one of them. Indeed, the whole Universe had to be affected at the same time, because the distance between entangled quanta did not matter. The "reality-forming" effects of human observations had to be described as non-local (*viz.*, "spooky action at a distance").

In spite of this apparent failure of "Einstein realism", it still followed that quantum mechanics was inconsistent with the Copenhagen interpretation. Suppose that two contradictory realities existed at the same time, until an act of measurement on particle A forced (non-locally) the entangled particle B into a well-defined sharp state. This second particle (B**)** would still be available for measurement. Therefore, it could be used to obtain a sharp observation for a conjugate variable, directly contradicting the principles of quantum mechanics. Bohr insisted that quantum mechanics was correct, as well as his interpretation, because of the game-changing role of complementarity ([22](#)). Still, he was not able to explain exactly how this process would work for entangled quanta. The debate reached a stalemate and remained undecided for a long time. Eventually, Bell's Inequality ([23](#)) inspired the development of theoretical and experimental tools for accurate verification ([24-27](#)). These advancements were able to prove that conjugate properties *can* be sharply determined at the same time, by virtue of entanglement. Yet, these observations did not contradict the formalism of quantum mechanics (and Heisenberg's principle, in particular), because Bell-type inequalities were conclusively violated. In short, the evidence suggested that quantum states were sharp and broad at the same time *after* the act of measurement. According to Bohr, the process of observation was supposed to force the Universe into a unique self-consistent reality, compatible with the manifestation of classical phenomena. Therefore, his interpretation was falsified. At the same time, the approach of Einstein and his supporters was equally discredited, because they expected the particle properties to be well-defined prior to measurement. No party was right, but quantum theory continued to work with unparalleled precision. The paradox remained unsolved.

## The Solution

The development of classical wave mechanics was influenced by two major approaches. One was the method of ray tracing (based on geometrical analysis). The other was the method of wavelet integration (based on the Huygens-Fresnel formalism) (*15*,*28*). By the end of the 19th century, the established practice was to assume that ray tracing captured the true ontology of wave propagation, in contrast to wavelet integration that entailed correct predictions "for the wrong reasons" (*29*). When the classical paradigm was contradicted by quantum mechanical observations, a new way of thinking was required. As shown above, Niels Bohr carried the day by reversing the status of "real" components and "illusory" measurement artifacts. Still, he preserved the method of ray tracing as the backbone for his new interpretation. In retrospect, a different solution was also possible. Namely, one could assume that wavelet integration captured the true ontology of propagation, while ray tracing was the method with correct predictions for the wrong reasons. The challenge is to define the ontological essence of the wavelet model in a practically relevant way.

Mechanical waves (such as sounds) can be described as chains of elastic interactions. As suggested above, a series of momentum-transfer events can be identical to the undisturbed motion of a free particle. For example, if a ball rolls through a tunnel, it is hard to guess if it went through undisturbed, or if it bumped into another identical ball, sending it forward. The second ball, in turn, could have also collided with another ball, and so on. This formal equivalence enables the use of virtual particles as model systems for real waves, with remarkable practical advantages for quantitative analysis. The problem is that 3-dimensional mechanical waves do not evolve through 1-dimensional interactions. Molecules that make up gases and liquids do not "bump" into each other like billiard balls. Instead, they have indiscriminate effects, pushing every particle in their vicinity with their fields. This effect is adequately captured by Huygens' concept of "wavelet", because the electromagnetic forces of each molecule obey the inverse square law and are most likely to have a hemi-spherical effect on the medium in the direction of their displacement. Thus, it is not possible to produce a compelling wave ontology in terms of rectilinear processes at the microscopic level. Unfortunately, this nuance is lost in the macroscopic analysis of wave propagation. Even with modern computers, it would be a daunting task to calculate the net effect of all the molecules on each other, in a volume of oscillation. Yet, the symmetry of Huygens-Fresnel propagation enables a methodological shortcut. Instead of calculating the shape of every intermediate wave-front between a source and a detector, it is possible to calculate the profile of the final wave-front directly. Every point in the plane of origin is associated with a macroscopic sphere (large wavelet), tangent on the plane of observation. The superposition of all the spheres (or, rather, their outer edges in the plane of observation) is used to calculate the observable net state at the destination (*30, 31*). It is important to emphasize: this shortcut works because it gives the same result as the tedious process of calculating large numbers of intermediate steps. However, it has become the only method used in practice and grew to be perceived as the correct way to interpret wave propagation. In other words, every point in the input wave-front is associated with a small number of relevant points in the output wave-front (where the tangent of the associated wavelet is consequential), and all of them can be visually connected with rays (*32*). As a result, Huygens-Fresnel propagation is actually interpreted with a "ray-tracing" model.

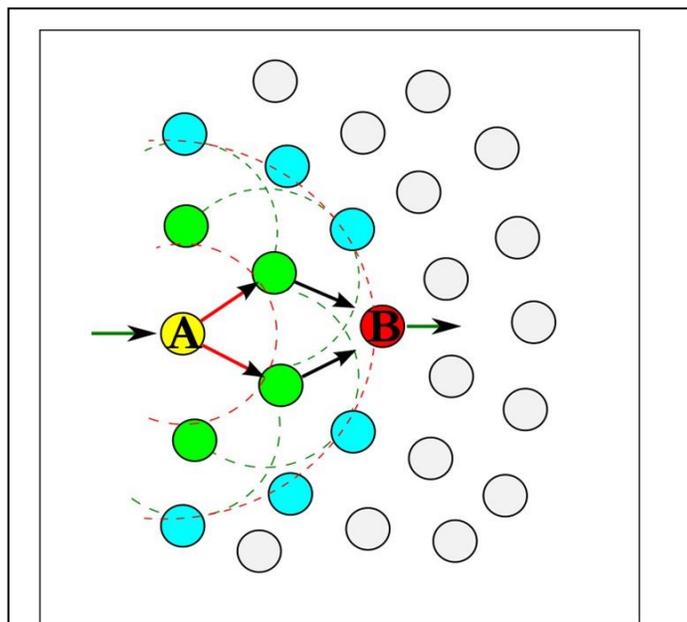

**Fig. 2. Suggested mechanism for wavelet propagation.** A displaced molecule (A) pushes all the adjacent particles with its field, producing a wavelet. All the displaced particles have a similar effect on their surroundings. Yet, many molecules at the receiving end (such as particle B) experience the effect of several forces. Classical objects cannot move in multiple directions at the same time. They can only move in the direction of the net force, producing wavelets accordingly. That is why classical mechanical waves cannot really go through each other unperturbed. Only the net state of superposition can be real.

From an ontological point of view, this practice is problematic on several levels. First of all, the large wavelet approximation only works for perfectly homogeneous media. If the process of propagation is constantly uniform and lossless, then the same final wave-front is predicted by any intermediate net state. The shortcut works, because the intervening stages are inconsequential. This is not the case for inhomogeneous media, where correct predictions can only be obtained by taking into the account the local structural details and by time-evolving the wave-front in small increments (*15, 31*). Hence, the solution with one-dimensional "independent" modes is not general enough to be ontologically compelling. Secondly, this approach entails conceptual difficulties. A wave is by definition a detectable oscillation on a medium. If any future state is predicted with separable components, then the whole space becomes "the wave", and the observable oscillation becomes "an artifact". For example, a coherent light beam passing through a lens produces an Airy pattern in the focal plane. This observation is typically explained by showing how various modes cancel out at the detector, and only a small region contains the effects of constructive interference. In other words, the modes are still assumed to pervade the whole medium of propagation, and the observable process of energy redistribution is treated as non-physical. "The wave" is no longer the wave.

Finally, this practice ignores the mechanical insight that motivated the formulation of Huygens' Principle. A ripple in a pond can be expressed as a collective effect of a row of wavelets. How does this actually work? Molecules are displaced and their fields impact every adjacent entity. Yet, every molecule at the receiving end is necessarily affected by many wavelets at the same time (Fig. 2). When classical objects experience different types of forces, they have to react by moving in the direction of the net vector. The components are still there, in a quantitative sense, but they are physically replaced by the equivalent single output. In other words, the wavelets are idealized patterns that appear spherical only in isolation. In actual conditions, the receiving molecules have to express collective effects directly. If the net state was not always real, then single objects would have to move in several directions at the same time, or even transfer momentum without moving at all (*33*). To sum up, the ontology of wavelet propagation is radically distorted by the assumption of separable modes. Even classical mechanical waves appear problematic in this light. The question is: do interpretive problems persist in models with *inseparable* components, in which the net state (and only the net state) is assumed to be real at every step?

A well-known problem in wave theory is the interpretation of coherent interference. What mechanism is hiding under the manifestation of fringe patterns? Typically, the nature of this process is explained by reference to amplitude superposition. At every point of observation, the net state is exactly equal to the vector sum of overlapping components. In some areas, the amplitudes happen to act in the same direction (peak to peak). In others, they work in opposite directions (peak to trough). This explanation requires undisturbed and separable input waves, or else – it seems – the fringes would not look the way they do. Unfortunately, the amplitudes are not directly proportional to the amount of incident radiation. Instead, they have to be squared for proper analysis. Hence, the amount of energy in the fringes is not a linear sum of input components. In particular, the center of a bright fringe in the Young interferometer contains twice as much power as the sum of individual components measured one by one (Fig. 3). Indeed, it contains the same amount of energy that is found to be missing in the areas with destructive interference. In a ray model of wave propagation, this redistribution of energy can only be treated as non-physical (an illusory artifact of measurement). Yet, the practical consequences of this process are always physical. It is not possible to extract useful energy from dark fringes, while the extra power is readily available in the bright fringes. Let us consider the same problem from a molecular point of view, in a classical context. Propagating waves always push forward the same total momentum, but sometimes particles act on each other in opposite directions, and they have to recoil. As a result, some volumes experience higher oscillations, at the expense of others. This approach entails that linear amplitude addition is a quantitative artifact of wave interference (not the other way around). As seen on geometric diagrams, the waves appear to go through each other. As seen in real life, interference takes place. The molecules in a medium do not travel with the waves. They can be associated with virtual wave-fronts that go through each other or that undergo specular reflection. Either assumption entails the same final predictions (*34*). What matters for the present discussion is that a wave – by definition – is the same thing as the motion of particles that make up the medium. It is not "something else" that pushes them around. Therefore, it is quite plausible (and even preferable) to interpret the observable net states of linear wave superposition as objective qualities at any level of analysis. If they are no longer treated as artifacts, quantum observations can be described as real, exactly as postulated by Niels Bohr, but without the need to suspend classical principles.

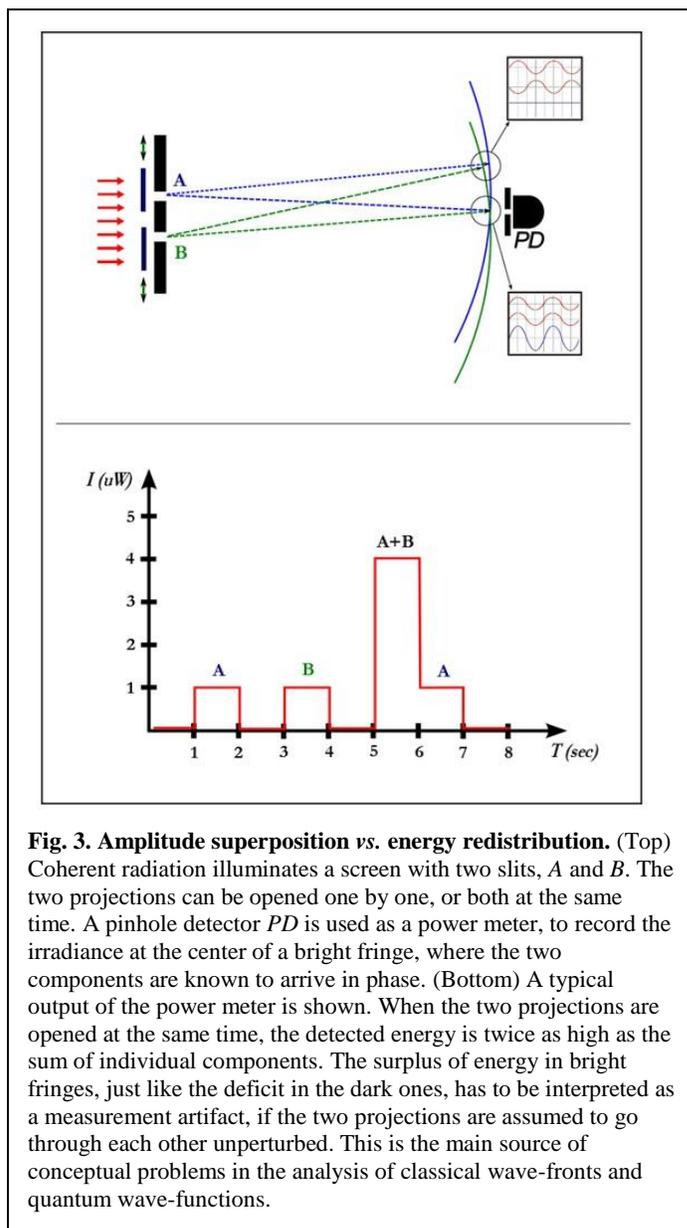

**Fig. 3. Amplitude superposition *vs*. energy redistribution.** (Top) Coherent radiation illuminates a screen with two slits, *A* and *B*. The two projections can be opened one by one, or both at the same time. A pinhole detector *PD* is used as a power meter, to record the irradiance at the center of a bright fringe, where the two components are known to arrive in phase. (Bottom) A typical output of the power meter is shown. When the two projections are opened at the same time, the detected energy is twice as high as the sum of individual components. The surplus of energy in bright fringes, just like the deficit in the dark ones, has to be interpreted as a measurement artifact, if the two projections are assumed to go through each other unperturbed. This is the main source of conceptual problems in the analysis of classical wave-fronts and quantum wave-functions.

Coherent optical beams produce exceptionally stable fringe patterns. Some regions are continuously dark, while others are continuously bright. The frequency of light is too high to notice phase relationships with the naked eye. Hence, it is too easy to ignore them. That is why it seems so hard to imagine that a fringe pattern can evolve into a split-beam projection (as if two beams intersected). For instance, interference gratings emit wave-fronts that are similar to fringe patterns, yet these projections do not reconfigure in the same way. If interference is real for beam superposition, how can it erase its own effects? The Huygens-Fresnel formalism gives a clear answer to this question: if the phases and the amplitudes of each point in the net state are taken faithfully into account, the evolution of the wave-front must result in double projections, provided the initial conditions are appropriate. Though – at the microscopic level – the model does not contain "beams" followed by "interference". There is only wavelet interference: first in the net shape of two approaching beams, then in the net shape of a single fringe pattern, then again in the net shape of two departing beams. Traditionally, this aspect is ignored, because input beams are easier time-evolved independently and superposed at arbitrary planes of observation. Yet, the method would contradict itself if two consecutive net states would not follow the same dynamics as the components that are presumed to generate them. (As a reminder, the original Huygens principle was not designed to calculate the net states of wavelet interference and made erroneous predictions. That is why it had to be corrected by Fresnel. Yet, once superposition is part of the model, it can be applied at any stage of propagation). This conclusion is naturally supported by considerations about classical media, in which only net states of motion are physically possible *(33)*. Though, it is also confirmed by optical experiments. For example, a double-slit fringe pattern can be recorded on a holographic plate. If such an imprint is illuminated with a *single* beam, it re-creates the same projection as the originally captured configuration. The complex phase details (unobservable in averaged patterns on a screen) are transferred in real time onto the emerging "hologram" *(35, 36)*. As a result, the projection is able to evolve into a split pattern (either directly, or through a lens, depending on the initial conditions), just like an actual two-beam superposition. Hence, the functional relevance of the net states of interference is not just a mathematical possibility – it is an empirical fact. Despite apprehensions to the contrary, classical wave mechanics is naturally compatible with the simultaneous reality of fringes and separable output beams. Therefore, it is not necessary to invoke exotic principles (such as complementarity) in order to explain their manifestation.

Another implication of this realist approach to Huygens-Fresnel propagation is the non-existence of constant input properties. At any given stage, the wavelets overlap to produce a new frontline. This construct, in turn, is a source of new wavelets that interfere to produce another frontline, and so on. (Indeed, the frontline is an approximation. For molecular media, field superposition operates at the level of each individual particle that is engaged by the travelling perturbation). Whenever a wave property has a sharp value, it happens because the individual contributions add up (and/or cancel out) to a spectrally narrow state. In plain language, the particles "just move". Their motion cannot contain real separable components. This conclusion is in obvious contrast to ray tracing models, where "input" values are presumed to be created at the source and carried unperturbed to the detectors *(37)*. Accordingly, a mode of propagation represented by a straight "ray" can have sharp values at one location and broad values at the next. Yet, this would not be due to the context of measurement. Instead, it would be a reflection of the objective microscopic qualities of the wave (ideally, captured by the act of measurement with fidelity). An interesting off-shoot of this conclusion is the possibility of a classical mechanism for Heisenberg's Principle. Conjugate variables are by definition Fourier transforms of each other *(9, 30)*. This means that one of them is forced to have a narrow spectrum, when the other is wide. A process of interference can only produce a net state in which one variable or the other is sharp in such a pair, but never both. In this interpretive context, broad spectra do not signal "lack of knowledge". They are natural states of wave variables. In order to observe sharp states, a wave has to be prepared accordingly, or captured at an appropriate location.

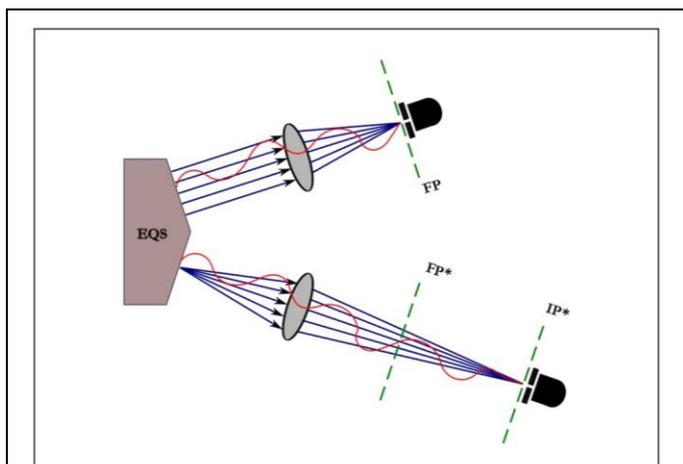

**Fig. 4. Quantum correlations without paradox.** A source *EQS* of entangled quanta (photons) produces two beams. Each photon in the upper beam is identical to a photon in the lower beam, by design. Therefore, the two beams are macroscopically and microscopically identical. Two identical photons will experience identical interactions, each in its own path. As a result, they will have identical trajectories (*red lines*) and will display identical qualities, if measured in the same way. Yet, their properties always correspond to the net state of the wave-function at the point of observation. In the focal planes (*FP* and *FP\**), each of them will have a sharp momentum and wide position spectrum, as if they belong to collimated beams emitted straight from the source (*upper set of blue rays*). In the image planes (*IP* not shown and IP\*), their spectra will have complementary profiles, as if they emerge from known points on the emitting surface (*lower set of blue rays*). In actuality, the photons belong to multi-mode wave-functions, and the shown geometric profiles have nothing to do with their actual history of propagation. The important aspect is that each photon obeys the restriction of Heisenberg's principle locally, in its own context, at every stage of propagation.

As a corollary of the above, the weirdness of quantum mechanics follows from the efforts to interpret the properties of the wave-function with particle models of wave propagation. As a population of (virtual) particles, the wave-function appears to carry a set of sharp states from the source to the detector. For this reason, broad states are perceived as "less real". They appear to contain a "washed-out" mix of input states that cannot be resolved by detectors. In contrast, a wavelet-based model suggests that all the states are equally real as wave-function properties, regardless of their spectral composition. The microscopic interference of wavelets is always a process with numerous components. The narrow states are just a special case in which these components generate a simple net profile. Yet, the same process of interference generates narrow and broad states at the same time, because many variables are conjugated. No less importantly, the properties of the wave-function can only be interpreted as dynamic and local. At every new frontline these attributes can change, depending on the context of propagation. This insight can be used to develop a paradox-free interpretation of entanglement.

Consider the simple example in which two wave-fronts are perfectly identical at emission, and detection events are selected with pinhole detectors from identical locations in the cross-section of each projection. If sharp values are desired for one variable, the data needs to be collected from a plane of observation where they are predicted to be possible. For example, momentum states have sharp spectra in the focal plane of a lens (Fig. 4). If both projections are processed with identical lenses, coincident events will have sharp and identical momentum profiles. At the same time, both of them will expose broad and identical position spectra. The broad position spectrum and the sharp momentum spectrum are necessary and inseparable properties of this context of observation. On the other hand, if sharp values for position are required, then observations need to be carried out in a different plane, *i.e.* the image plane of the same lens. Necessarily, the spectra of momentum states will be maximally wide in this new volume of detection. To sum up, identical measurements will produce identical results in the two projections, but no "spooky action at a distance" is suspected in this case. Furthermore, each context of observation has its own independent profile. The sharp position state from the image plane and the sharp momentum state from the focal plane cannot be attributed to "the same" entity, because both of them are *local* products of linear superposition between multiple interfering components. Granted, it is possible to determine two sharp values for two non-commuting variables of the same entity, by recording coincidences between the two projections in different planes of observation. Still, these events do not belong to compatible populations, because they come from different contexts. In contradistinction with previous models, the two sharp values cannot be described as *simultaneous* hidden properties of the same entity. Therefore, the EPR paradox is avoided. To be clear, the problem is not removed by showing that conjugate variables are sharp in different contexts, because the same observation is present in other approaches. It is solved by interpreting the sharp states as byproducts of dynamic superposition (*i.e.*, local net states of wavelet interference). It does not matter if these observations are made with classical beams or single quanta, because wave-function properties are expressed in both cases.

## Discussion

The main property that appears to force the split between quantum mechanics and classical mechanics is *separability (38, 39)*. When a classical particle system produces observable events, it is possible to detect macroscopic joint effects with inseparable components. However, particles do not cease to exist just because they act together. At least as a matter of principle, their individual behavior is theoretically and empirically separable at the microscopic level. In the early days of quantum mechanics, both Einstein and Bohr expected elementary particles to display separability. Einstein changed his mind after studying the work of Bose, while Bohr switched positions after the disproof of the Bohr-Kramers-Slater theory *(40)*. By the time of the 1927 Solvay meeting, both of them agreed that quantum states should be inseparable (*i.e.*, entangled) and both of them perceived this as a radical departure from "realist" classical physics. As it is known, Bohr saw this as a sign that new concepts are needed for the understanding of the Universe (such as complementarity), while Einstein worried that quantum mechanics cannot capture the complete reality of microscopic interactions. Further on, this insight informed Schrodinger's work on his famous wave equation *(41)*. The latter was purposefully developed in 3N-dimensional phase space, in order to express the primary relevance of the net states of linear superposition between inseparable components *(40, 42, 43)*.

The concept of separability has grown into a fountain of confusion over time, because of the difficulty of reaching consensus over its object. What exactly is inseparable in quantum mechanics? As the word "inseparable" was replaced by the word "entangled", and as the debates increasingly focused on experiments with correlated entities, the widespread perception has become that different *events* can become inseparable from each other. Yet, this is obviously incorrect, because correlated systems are detected in separate conditions (one "over here", and the other "over there"). By definition, "entangled" quanta emerge after *breaking apart* from a common system and are produced in a manner that enables independent observations. The real issue in this context is the nature of a particle that seems to be "in many states at the same time". When a single object expresses the effect of multiple components, it can only move in the direction of the net state. Therefore, the many constituents of a wide spectrum cannot be detected one by one in actual experiments. Einstein and Bohr had been sparring over separability long before the EPR paper *(44-46)*. They both saw it as a measurement artifact, and disagreed over the ontological significance of the act of observation. Indeed, correlated quanta were brought into this discussion in order to explore the possibility of undetectable separability *(47-49)*. A particle must be assumed to have many states at the same time prior to measurement, yet the reality of individual components can be exposed indirectly, by measuring a correlated "twin" (or, so it seemed at the time). Another way to look at this is that a quantum can be measured only once, yet more than one property can be determined with the help of correlated systems. Hence, the object of correlation tests is to determine the relationships between multiple qualities of a *single* quantum. If one of these qualities has inseparable components, then the relationships between them will obey certain rules (*e.g.*, Heisenberg's Principle, whose effect determines the profile of Tsirelson's Inequality). The true mystery of quantum mechanics is that single particles express inseparable qualities when detected one at a time, and joint effects seem impossible.

As shown above, the special concepts and principles of quantum mechanics can be derived from the details of microscopic interactions that determine the properties of classical mechanical waves. The crucial insight is that separability applies to the joint effects of *discrete particles*. When it comes to the nature of individual *states of motion*, there are no logical or empirical reasons to invoke it. Yet, waves require an ontological explanation in terms of states of motion (for which the net states alone can be real), because the medium does not travel with the propagating oscillations. Accordingly, the properties of quanta cannot be expected to be necessarily separable when determined by their context of propagation. Incidentally, this conclusion does not depend on any particular assumption about the nature of quantum wave-functions. As long as the properties of quanta are determined by an external process with wave-like net states, the same considerations apply. Likewise, this presentation does not entail that waves alone are real and that quantum particles do not exist. The only necessary implication is that event-producing quanta (regardless of their actual nature) must express the qualities of the associated (or guiding) wave-functions. It is not the particles that add up to the profile of a wave-function. It is the wave-function that determines their individual properties. To some readers, such conclusions might sound counterintuitive. After all, so many textbooks insist that quantum mechanics is irreducibly non-classical. The thing to keep in mind is that classical wave interactions have been interpreted with particle models (derived from ray tracing), while classical particles are – of course – also interpreted with particle models. For this reason, separability appears to be a necessary component of any kind of classical approach. Still, the main point of the preceding argument is that classical wave models can be developed in two different ways, and that one of them is free of such conceptual complications.

The obvious question is to ask is: what about non-locality? Let us recall why it was seen as necessary in the first place. When the momentum of a quantum is determined with a sharp measurement, the state of its correlated partner is also known. If this analysis is performed within a particle model of the wave-function, then the inferred quantity is assumed to persist with the quantum from the source to the detector. Therefore, this sharp momentum state is assumed to be present even when the input position of the second quantum is "revealed". Consequently, correlated particles are expected to have sharp states for momentum and position at the same time, and must also produce correlation coefficients below the limit of any Bell-type inequality *(2)*. This interpretive conclusion is at odds with quantum theory and the experimental record. Therefore, some sort of invisible demon must be invoked for compliance with the established facts. The name of this invisible demon is *nonlocality*. Hence, "spooky action at a distance" is just an interpretive element, intended to reconcile a contingent assumption with a verifiable conclusion. Its only purpose is to explain why two variables with apparently simultaneous sharp states do not commute. On the other hand, if the wave-function properties are interpreted with a wavelet model, then the ontological implications are fundamentally different. When momentum and position are macroscopic wave properties, they never commute. They represent collective effects, and their spectra – by definition – can only be inversely proportional at every step of propagation. As shown above, the process of sharp observation of two conjugate variables requires mutually exclusive contexts of propagation. Accordingly, they still obey Heisenberg's principle locally. If so, then entangled quanta can always display correlations that reach as high as Tsirelson's limit *(6, 8, 50)*. Simply put, predicted violations of Bell-type inequalities are due to the fact that wave-function variables are wave (rather than particle) properties. In the absence of any conflict between interpretation and theory, there is no need to invoke invisible demons.

## Summary

The macroscopic properties of classical waves can be predicted with two equivalent models. On the one hand, they can be analyzed as a stream of (virtual) particles with rectilinear trajectories that go through each other unperturbed. On the other hand, they can be explained through a complex process of wavelet interaction. In the first case, wave superposition is perceived as an artifact of measurement, because detectors are presumed to respond to multiple independent components at the same time. In the second case, superposition is an objective process that works at the microscopic level. These two models entail the same macroscopic predictions for a large class of phenomena, but their microscopic unobservable implications are radically different. In particular, the particle model is not able to explain the nature of wave diffraction, or the relevance of frequency for the parameters of interference. More importantly, it leads to fundamental conceptual inconsistencies during the interpretation of all types of waves, from mechanical oscillations to quantum wave-functions. These complications are absent in the wavelet model. Therefore, the puzzles of quantum mechanics, including the EPR paradox, can be solved by switching the interpretation of wave-function properties from a particle to a wavelet model.

Quantum phenomena are observed by accumulating large numbers of discrete events. The most intuitive explanation is that individual events are produced by objectively discrete entities (*i.e.*, quanta). Yet, the distributions of these events cannot be interpreted as patterns of "underlying" particle behavior. They always reflect wave properties, which is why they are predicted with notorious accuracy by wave-functions. In short, scientists detect particles, but acquire information about wave properties. A special quality of all the waves is their ability to express the effect of multiple states in superposition. If the mechanism of wave propagation is interpreted with a particle model, all the superposed component states are assumed to be independent and real at the same time. Indeed they are presumed to be carried intact from the point of emission to the point of detection. In contrast, if the mechanism of wave propagation is interpreted with a wavelet model, the superposed components are treated as virtual. Different wavelets from different source points interfere, producing a net state. From a physical point of view, only the net state is real in this analytical context. The so-called components are derived by breaking down the observed complex state into simple constituents (Fourier decomposition). They represent the minimal list of virtual ingredients that could be used to synthesize the net state in the easiest way possible. If most of the actual (wavelet) components cancel out and only a simple net state remains, then the output spectrum is sharp. If the number of such components remains large, because of the nature of phase relationships between real inputs, then the net spectrum is wide. In short, particle models of the quantum wave-function describe permanent properties, distorted by the context of measurement. Wavelet models describe transient local properties, created by the context of propagation. In this approach, if quanta are assumed to be real, they must always switch to the net state of their local environment, like chameleons. At every stage of propagation, their properties can change, but they are always captured by the act of measurement in their objective state (assuming no measurement artifacts).

The essence of the EPR paradox is the implication that quanta have broad states even after exhibiting sharp states in special measurement settings, as if their spectra are broad and sharp at the same time. This problem is caused by the assumption that quanta express permanent particle qualities. When a measurement reveals a narrow spectrum, it is assumed to capture the "real" state (which is otherwise smeared by "low resolution" observations). In this approach, two entangled quanta are assumed to be correlated because of the relationships between their intrinsic properties, regardless of their context of propagation. Hence, when a sharp property is captured in one context and the second one is recorded in another, both properties are attributed to the same entity at the same time, with paradoxical consequences. None of these complications arise in a wavelet approach in which quantum states are interpreted as transient and local. If the net states of superposition are always real, then the same level of reality is attributed to a quantum property whether its spectrum is sharp or broad. More importantly, when a quantum is measured, the state of its entangled partner is only determined for an identical context of propagation, where it can be confirmed by experiment. In all other contexts, the quanta are assumed to have different properties, with different spectral profiles. The correlation between entangled partners follows from the properties of their contexts of propagation – they are presumed to experience identical (or otherwise correlated) stages of interference at each step, after breaking apart from a joint quantum system. Moreover, conjugated variables have spectra with inversely proportional width in each context of propagation, simply because they represent wave qualities. Thus, quanta cannot be described as having several sharp states at the same time for non-commuting features, and the EPR paradox is avoided.

This solution was obtained by reinterpreting the sharp quantum spectra as outcomes of wavelet interference. Previously, the apparent problem was to explain the nature of states with wide spectra. ("How is it possible for a quantum to be in many states at the same time?") The sharp states were assumed to be real by default. In contrast, the new approach entails that simple quantum states are problematic. ("How is it possible for a quantum property to have a sharp spectrum, if every state is a result of multiple component superposition?") The explanation is that narrow spectra emerge in special cases, when multiple inputs interfere and cancel out (or add up) to a simple net state. When such qualities persist in space and time, it is because they are continuously (re)created by waves, rather than transported unperturbed by particles.

**Acknowledgments:** The content and structure of this manuscript was heavily influenced by the comments and insights of Ligia Mardari on earlier drafts.